\documentclass{article}
\usepackage{spconf,amsmath,graphicx}
\usepackage{multirow}
\usepackage[colorlinks,linkcolor=blue]{hyperref}

\title{Remove Appearance Shift for Ultrasound Image Segmentation via \\Fast and Universal Style Transfer}
\name{\parbox{\linewidth}{\centering Zhendong Liu$^{1,2,\dagger}$, Xin Yang$^{1,2,\dagger}$, Rui Gao$^{1,2}$, Shengfeng Liu$^{1,2}$, Haoran Dou$^{1,2}$, Shuangchi He$^{1,2}$, \textit{Yuhao Huang$^{1,2}$, Yankai Huang$^{3}$, Huanjia Luo$^{3}$, Yuanji Zhang$^{3}$, Yi Xiong$^{3}$, Dong Ni$^{1,2,*}$}} \thanks{$\dagger$ Authors contributed equally.} \thanks{* Corresponding author: \textit{nidong@szu.edu.cn}.} \thanks{This work was supported by the grant from National Key R\&D Program of China (No. 2019YFC0118300); Shenzhen Peacock Plan (No. KQTD2016053112051497, KQJSCX20180328095606003); Medical Scientific Research Foundation of Guangdong Province, China (No. B2018031).}}

\address{$^{1}$National-Regional Key Technology Engineering Laboratory for Medical Ultrasound, \\Guangdong Key Laboratory for Biomedical Measurements and Ultrasound Imaging, \\School of Biomedical Engineering, Health Science Center, Shenzhen University, Shenzhen, China \\$^{2}$Medical UltraSound Image Computing (MUSIC) Lab, Shenzhen University, Shenzhen, China\\$^{3}$Department of Ultrasound, Luohu People's Hosptial, Shenzhen, China}

\begin{document}
%
\maketitle
\begin{abstract}
Deep Neural Networks (DNNs) suffer from the performance degradation when image appearance shift occurs, especially in ultrasound (US) image segmentation. In this paper, we propose a novel and intuitive framework to remove the appearance shift, and hence improve the generalization ability of DNNs. Our work has three highlights. First, we follow the spirit of universal style transfer to remove appearance shifts, which was not explored before for US images. Without sacrificing image structure details, it enables the arbitrary style-content transfer. Second, accelerated with Adaptive Instance Normalization block, our framework achieved real-time speed required in the clinical US scanning. Third, an efficient and effective style image selection strategy is proposed to ensure the target-style US image and testing content US image properly match each other. Experiments on two large US datasets demonstrate that our methods are superior to state-of-the-art methods on making DNNs robust against various appearance shifts. \par
\end{abstract}
\section{Introduction}
In recent years, Deep Neural Networks (DNNs) have dominated the field of medical image analysis \cite{LIU2019261}. However, DNNs have the performance degradation when image appearance shift exists between the model training and testing phases. Therefore, DNNs cannot guarantee the accuracy in real clinical scenarios \cite{stead2018clinical_jama,yang2018generalizing}. \par

\begin{figure}[htb]
	\centering
	\includegraphics[width=1.0\linewidth]{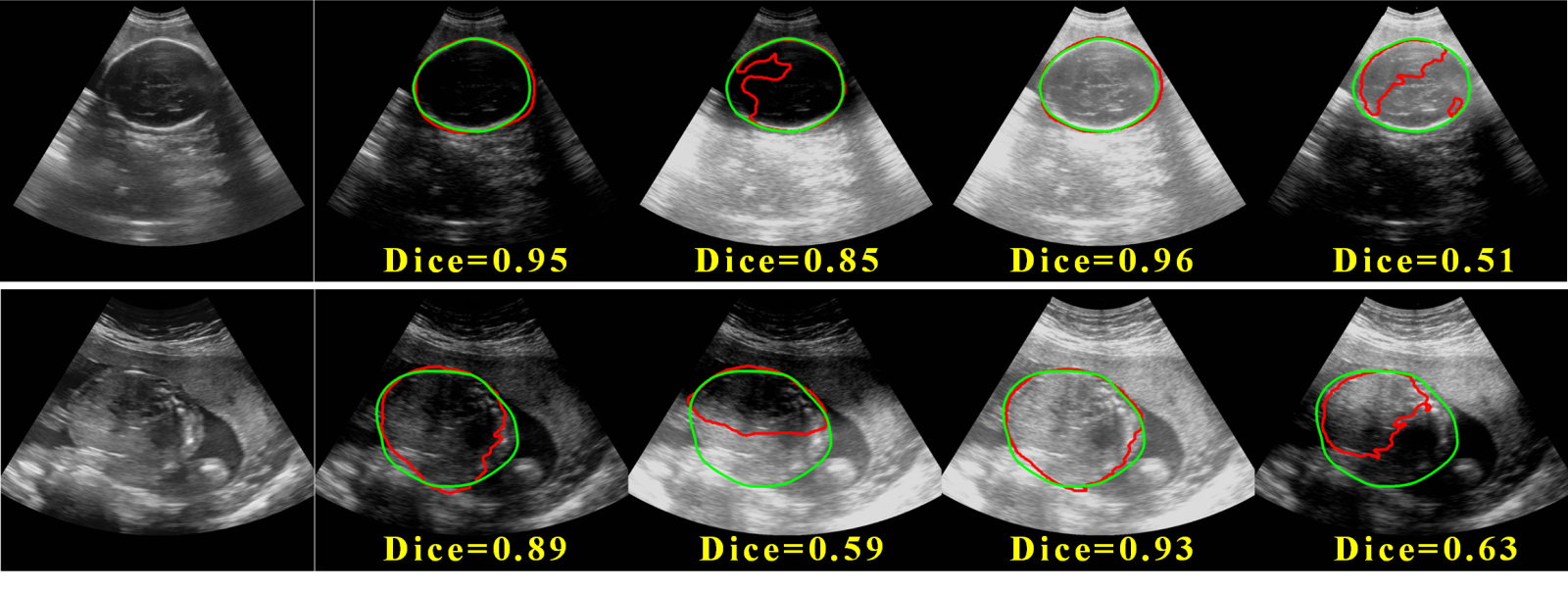}
	\caption{Original US images and four variants with different TGC. Two groups US images for the fetal head (top) and fetal abdomen (bottom). Yellow digits denote the Dice. Green and red curves represent ground truth and the segmentation results, respectively.}
	\label{fig:problem-task}
\end{figure}

The situation becomes more severe for DNNs based ultrasound (US) image segmentation. As shown in Fig.\ref{fig:problem-task}, there are two groups of US images from the prenatal scan. With different Time Gain Compensation (TGC) settings, tissues intensity in different depths are affected. The original US image changes into variants with obvious and typical appearance shift. A well-trained DNN then presents dramatically different segmentation results on the variants. Therefore, making DNNs robust against appearance shift on US images is highly desired. However, this task is non-trivial. First, DNN models in clinical US analysis are facing an open scenario, where DNNs only have access to limited source domain images/labels, and are blinded to target domain samples. Second, varitions in acoustic attenuation, scan operators, US machines, probes and empirical imaging parameters often make appearance shift unpredictable and hard to model. \par

Recently, Domain Adaptation (DA) was frequently used to remove image appearance shift. Kamnitsas et al. \cite{kamnitsas2017unsupervised} proposed to align domain features, but it needs a large number of images and labels from the target domain, which is infeasible in clinic scenario. Based on Cycle-GAN \cite{zhu2017unpaired}, translating the image appearance among domains \cite{huo2017adversarial} and shape guided image translation \cite{chen2018semantic} were proposed. However, GAN-based methods often introduce artifacts into the original image and make the image distorted. In addition, DA often limits itself among a fixed number of domains with limited appearance shift. This is not suitable for the open scenario with unpredictable image appearance shift. By revisiting the style transfer \cite{gatys2016image}, Ma et al. built an online scheme to remove appearance shift in the cardiovascular magnetic resonance (MR) image segmentation \cite{ma2019neural}. However, their style transfer method is unstable and may destroy image details. In \cite{kolkin2019style}, a new style transfer method, called STROTSS, was proposed to preserve image details for any style-content pairs during transfer. STROTSS is promising but it is time-consuming. It also outputs random results due to the resampling of hypercolumns. \par

In this paper, on the basis of Wavelet Corrected Transfer network (WaveCT) \cite{yoo2019photorealistic} and Adaptive Instance Normalization (AdaIN) \cite{huang2017arbitrary}, we propose a novel and general style transfer framework to remove US image appearance shift (denoted as \textit{WaveCT-AIN}). Our framework has three highlights: (a) It is the first work to explore universal style transfer for US image segmentation. WaveCT-AIN preserves the image structure details and enables arbitrary style-content transfer between images. Unpredictable US image appearance shifts are consistently tackled under this framework. (b) Compared to STROTSS, our WaveCT-AIN is lightweight and satisfies the real-time requirements. With AdaIN block for acceleration, it only takes 0.07s for an inference. (c) An efficient and effective style image selection strategy is also adopted to retrieve suitable and appearance-invariant US style image for optimized style transfer. Experiments on two US datasets demonstrate that the proposed WaveCT-AIN obtained a superior performance than the state-of-the-art methods. \par

\section{Methodology}
\label{sec:format}
\begin{figure}[htb]
	\centering
	\includegraphics[width=1.0\linewidth]{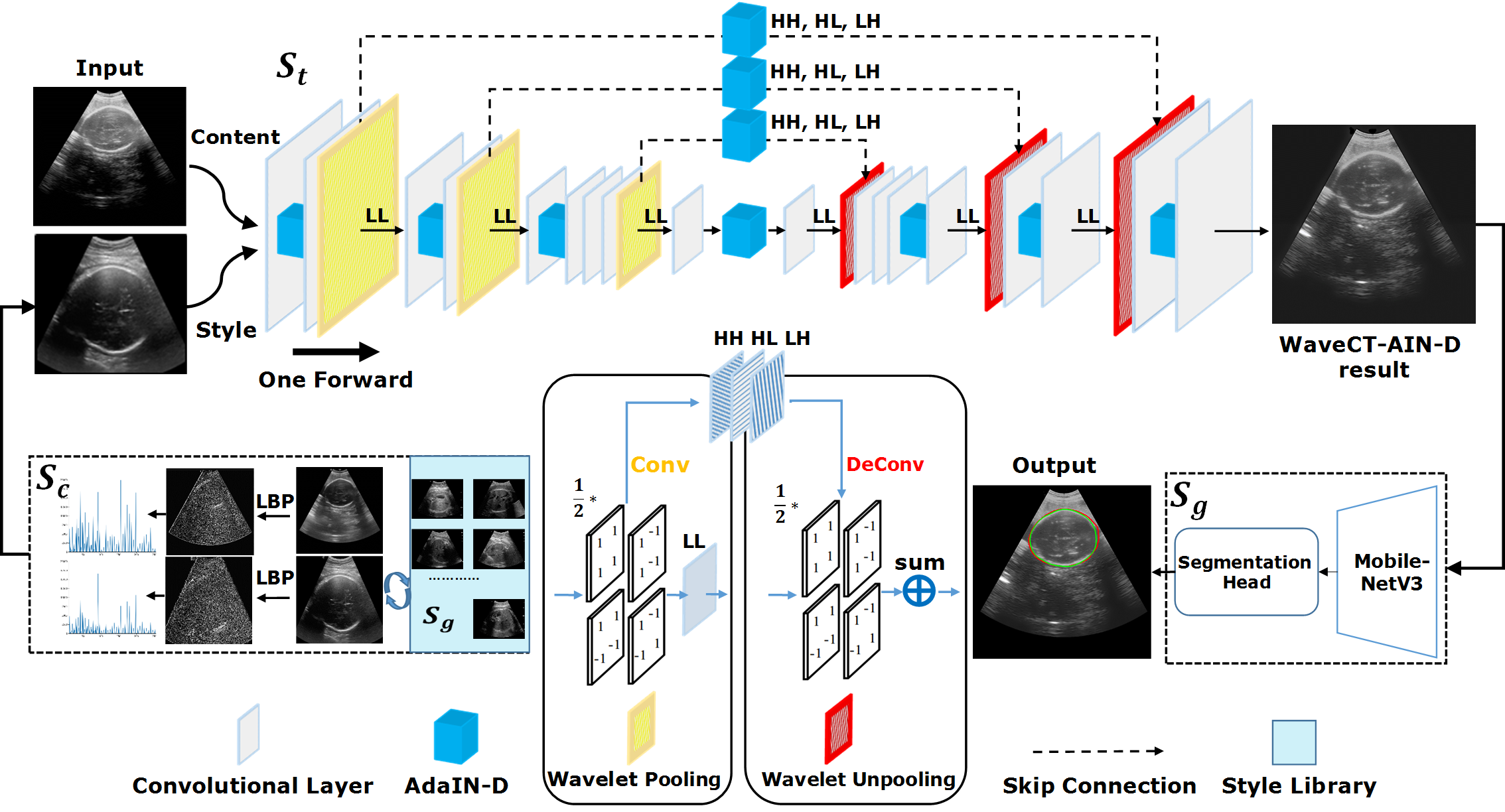}
	\caption{Overview of the proposed framework.}
	\label{fig:framework}
\end{figure}

The proposed framework is shown in Fig. \ref{fig:framework}. It consists of a style selection module $S_c$, a style transfer module $S_t$ (WaveCT-AIN-D) and a segmentor $S_g$. For a testing content image, $S_c$ retrieves an US style image from source domain as a reference and inputs it to $S_t$. Then, $S_t$ instantly removes the appearance shift in content image and makes it suitable for $S_g$ to segment. \par
\subsection{Universal and High Quality Style Transfer}
Style transfer can remove appearance shift to approach robust US image analysis. To meet clinical US scanning requirements, following aspects should be considered: (a) The method should enable \textit{universal} transfer between any style-content image pairs. (b) Transfer results should be \textit{stable} rather than fluctuant \cite{ma2019neural,kolkin2019style}. (c) Image structure details should be highly preserved. (d) Transfer process should be \textit{fast} to avoid high latency\cite{kolkin2019style}. (e) The method should provide \textit{style selection} strategy to provide case-specific style transfer. \par

In order to meet all the above needs, we adopt the WaveCT network recently used in WCT$^2$ \cite{yoo2019photorealistic}. As shown in Fig. \ref{fig:framework}, style image and content image are hierarchically processed by the encoder-decoder network in WaveCT. At several sites, multi-scale style information is distilled from the feature maps of style image. Style information is then transferred to modulate the feature maps of content image by style transform block. With the modulated features, decoder reconstructs the stylized content image. \par

Different from previous online style transfer methods which may distort image details \cite{ma2019neural}, the proposed WaveCT has two main advantages. \textit{First}, WaveCT replaces vanilla max-pooling/unpooling layers with the Haar wavelet pooling/unpooling layers, which can exactly reconstruct the stylized image with minimal structure loss. \textit{Second}, WaveCT splits the features into low-frequency and high-frequency components via Haar wavelet pooling. The low-frequency captures smooth textures while the high-frequency extracts edge-like features. WaveCT only passes the low-frequency in the main network and skips the high-frequency to the decoder. Due to this layout, WaveCT with style transform block can achieve universal and high quality stylization as we expected. \par

\subsection{Accelerating the Style Transfer}
WaveCT associated with the style transform block, i.e., whitening and coloring transforms (WCT) \cite{li2017universal}, is the work-\\horse of WCT$^2$\cite{yoo2019photorealistic}. Although effective, WCT$^2$ has high computation latency. Specifically, WCT firstly applies whitening transformation to erase the style information in the content feature map, and then a coloring transformation to render the texture of style image to the whitening result. Since these two steps involve heavy computation among multiple large matrices, WCT$^2$ needs about 4.58 seconds for an inference, which is not acceptable in US image analyses. \par

\begin{figure}[htb]
\centering
\includegraphics[width=1.0\linewidth]{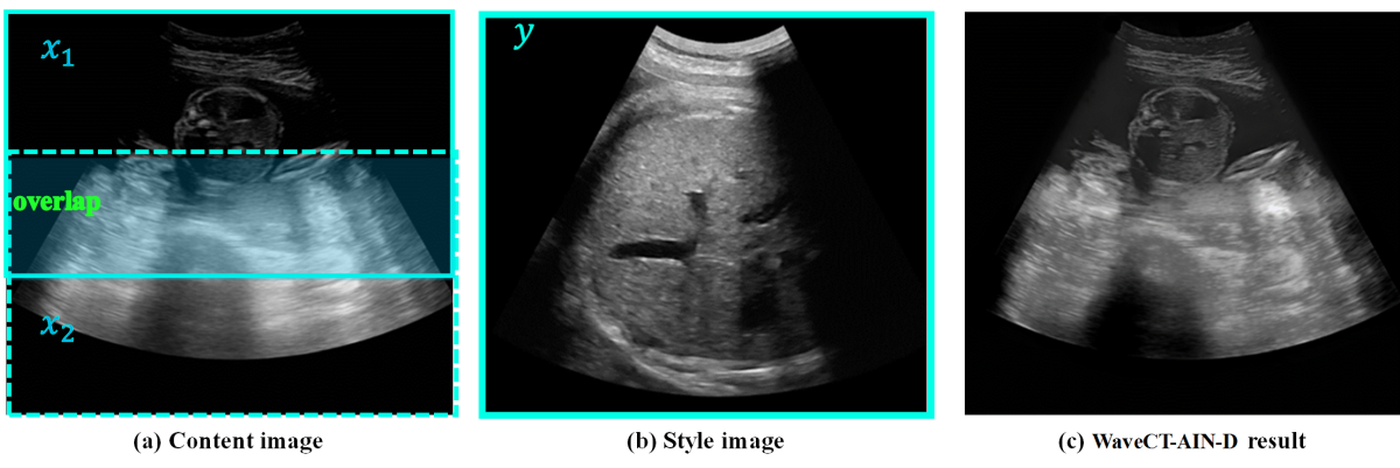}
\caption{(a) A content image split into two regions $x_{1}$ and $x_{2}$ with a half overlap. (b) A style image $y$. (c) Our WaveCT-AIN-D result.}
\label{fig:depth}
\end{figure}

To accelerate the transfer, we propose to upgrade all the WCT block in WCT$^2$ to a new block, AdaIN \cite{huang2017arbitrary}. The modified system, i.e., WaveCT with AdaIn, is denoted as \textit{WaveCT-AIN}. AdaIN is a variant of Instance Normalization. It only needs to align the channel-wise mean $\mu$ and standard deviation $\sigma$ of content feature maps $x$ to match those of target-style feature maps $y$:

\begin{equation}
AdaIN(x,y)=\sigma(y)\left(\frac{x-\mu(x)}{\sigma(x)}\right)+\mu(y)
\label{equ:3}
\end{equation}
AdaIN is lightweight, which significantly reduces the computation cost. Considering the fact that TGC changing along the depth is the main reason for the appearance shift in US image, we further modify the AdaIN into a depth-encoded version, \textit{AdaIN-D}. Specially, as illustrated in Fig.\ref{fig:depth}, where a window sliding along the depth is employed (Fig.\ref{fig:depth}(a)). The bandwidth and stride are set to two-thirds and one third height of content image respectively. Then we perform AdaIN transform for each region $x_i$ (i=1,2) and average the overlap. WaveCT with AdaIN-D is denoted as WaveCT-AIN-D (see Fig.\ref{fig:framework}). \par

\subsection{Case-specific Style Image Selection}
Due to the unpredictable appearance shift in testing images, style image should be adaptively determined to provide case-specific guidance. To achieve this, we propose a Local Binary Patterns (LBP) feature \cite{ojala2002multiresolution} based histogram matching strategy to improve the selection of style image. LBP is superior in capturing image textures and invariant to the changes in grayscale and rotation. \par

Based on the style image library from the source domain, the proposed strategy includes three steps. First, we obtain the histograms of testing content image $c$ and all the style images in the library according to the corresponding LBP feature spectrum. Second, the top-10 relevant style images are retrieved via the LBP histogram correlation. Third, to further eliminate the gap between the testing image and the retrieved style images, we localize the final style image with the smallest Euclidean distance. \par

Since AdaIN is sensitive to the mean and variance of style image, in the third step above, we require the mean and variance of the target style image $S_{T}$ to be as close as possible to those of content image $c$ instead.
\begin{equation}
S_{T} = \mathop{\arg\min}_{s_i\in\{s_1,...,s_{10}\}}{\Vert\mu(s_{i})-\mu(c)\Vert}_2+{\Vert\sigma(s_{i})-\sigma(c)\Vert}_2
\label{equ:0}
\end{equation}
where $s_{i}$ is each retrieved style image from the second step. $\mu$ and $\sigma$ are the mean and variance of the whole image.

\begin{figure}[htb]
	\centering
	\includegraphics[width=1.0\linewidth]{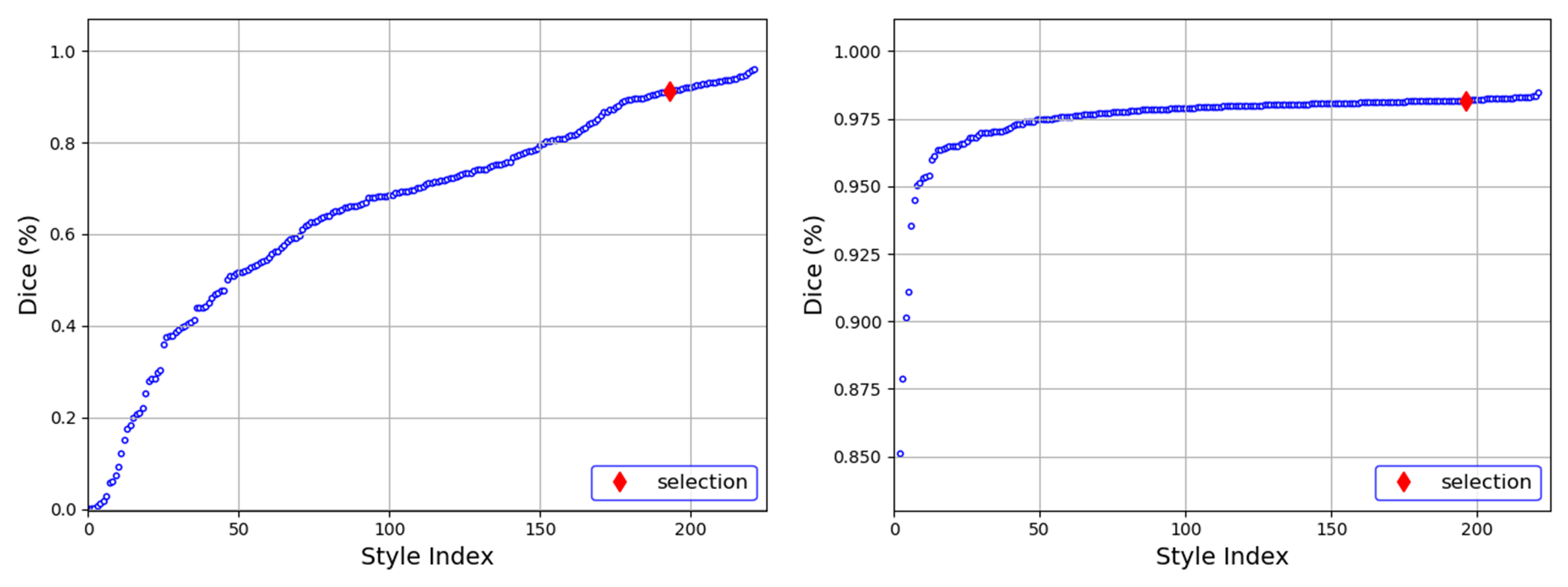}
	\caption{Our style selection method in (a) Gatys et al.\cite{gatys2016image} and (b) WaveCT-AIN-D for stylization results of a content image. Y-axis is the sorted Dice with corresponding style image index.}
	\label{fig:sc}
\end{figure}

Our proposed style selection strategy is effective and not only applicable to our WaveCT-AIN-D, but also to the classic method in \cite{gatys2016image}. In Fig. \ref{fig:sc}, we show the sorted segmentation results on an testing content image which is transferred from 222 style library images. Among all the results, our strategy hits a proper style image and generates superior segmentation results (red diamond), which indicates that our selected style can properly help remove the appearance shift. \par

\section{Experimental Results}
\subsection{Datasets and Implementation Details}
\textbf{Datasets.} Our methods were tested on 180 groups of fetal head (FH) and 198 groups of fetal abdomen (FA) US images. Each group consists of one original US image and four variants with different TGC (see Fig.\ref{fig:problem-task}). Approved by local IRB, all the images were anonymized and obtained by experienced experts using a Siemens system. Testing images cover the gestational age from 20 to 30 weeks. An expert provided delineation ground truth for all the images. Furthermore, 222 FH and 200 FA US images acquired by a GE scanner serve as the style image library. \par

\textbf{Implementation Details.}
We freeze the encoder of  WaveCT-AIN-D with VGG encoder weights. The decoder is firstly pre-trained on Microsoft COCO dataset and then fine-tuned with 3,000 US images, minimizing the L2 reconstruction loss. We use Adam optimizer with learning rate $10^{-3}$. In the style selection, we use LBP with uniform pattern and set the neighbour points to 8, the radius to 3. For the segmentation model, we choose the MobileNetV3-Large with a segmentation head as our segmentor backbone. Adam with learning rate $10^{-4}$ is adopted to train segmentor using an augmented source domain training set of 10,000 US images. Segmentor is pre-trained on the source domain without TGC change and frozen during testing. 
\par

\subsection{Results}
The quantitative comparisons between WaveCT-AIN-D and other methods on FA and FH ultrasound (US) images are summarized in Table \ref{tab:comparison_ac} and \ref{tab:comparison_hc}, respectively. Specifically, ``SS" denotes the method with our style selection, ``fix" denotes the method evaluated on the average of all the style library images based style transfer results. \par

\begin{table}[htbp]
  \centering
  \tiny
  \caption{Quantitative comparison on FA US segmentation. }
    \hspace*{\fill} \\
    \renewcommand\tabcolsep{4pt}
    \begin{tabular}{c|ccccc|c}
    \hline
    \textbf{Methods} & \textbf{Dice}(\%) & \textbf{Hdb}(pixel) & \textbf{Jaccard}(\%) & \textbf{SSIM}(\%) & \textbf{PSNR} & \textbf{Time}(s) \\
    \hline
    No processing & 85.60 & 21.83 & 79.27 & /   & /   & / \\
    HE  & 80.20 & 27.26 & 73.01 & 59.21 & 11.20 & 0.002 \\
    Cyclegan\cite{zhu2017unpaired} & 87.37 & 15.42 & 82.51 & 36.70 & 21.21 & 0.003 \\
    STROTSS\cite{kolkin2019style} & 88.71$\pm$1.22 & 15.34$\pm$1.88 & 83.21$\pm$0.83 & 41.34$\pm$2.55 & 20.26$\pm$1.41 & 181 \\
    Gatys et.al\cite{gatys2016image}(fix) & 70.13$\pm$2.31 & 42.32$\pm$3.50 & 65.56$\pm$1.42 & 24.23$\pm$3.30 & 16.10$\pm$2.03 & 3.19 \\
    Gatys et.al(SS) & 88.62$\pm$0.87 & 15.32$\pm$1.55 & 83.25$\pm$0.51 & 62.50$\pm$1.42 & 28.90$\pm$1.20 & 3.20 \\
    WCT$^2$\cite{yoo2019photorealistic}(SS) & 90.50 & 14.21 & 83.79 & 74.51 & 25.42 & 4.58 \\
    \hline
    \textbf{WaveCT-AIN(SS)} & 93.42 & \textbf{7.94} & 89.44 & 78.55 & 28.23 & \textbf{0.07} \\
    \textbf{WaveCT-AIN-D(fix)} & 92.34 & 9.83 & 88.52 & 79.43 & 29.34 & 0.10 \\
    \textbf{WaveCT-AIN-D(SS)} & \textbf{94.05} & 8.32 & \textbf{90.09} & \textbf{79.81} & \textbf{30.63} & 0.11 \\
    \hline
    \end{tabular}%
  \label{tab:comparison_ac}%
\end{table}%
\vspace{-0.6cm}
\begin{table}[htbp]
  \centering
  \tiny
  \caption{Quantitative comparisons on FH US segmentation.}
    \hspace*{\fill} \\
    \renewcommand\tabcolsep{4pt}
    \begin{tabular}{c|ccccc|c}
    \hline
    \textbf{Methods} & \textbf{Dice}(\%) & \textbf{Hdb}(pixel) & \textbf{Jaccard}(\%) & \textbf{SSIM}(\%) & \textbf{PSNR} & \textbf{Time}(s) \\
    \hline
    No processing & 91.10 & 16.28 & 86.39 & /   & /   & / \\
    HE & 85.03 & 20.44 & 80.49 & 59.15 & 11.04 & 0.002 \\
    Cyclegan\cite{zhu2017unpaired} & 93.36 & 9.02 & 90.05 & 36.66 & 21.14 & 0.003 \\
    STROTSS\cite{kolkin2019style} & 92.53$\pm$0.83 & 10.75$\pm$1.61 & 87.79$\pm$0.39 & 40.94$\pm$2.86 & 19.88$\pm$1.58 & 179 \\
    Gatys et.al\cite{gatys2016image}(fix) & 82.17$\pm$1.52 & 30.02$\pm$2.69 & 76.08$\pm$1.02 & 24.45$\pm$3.10 & 16.15$\pm$2.04 & 3.18 \\
    Gatys et.al(SS) & 93.55$\pm$0.35 & 12.51$\pm$1.76 & 91.57$\pm$0.22 & 61.30$\pm$1.04 & 28.70$\pm$1.12 & 3.19 \\
    WCT$^2$\cite{yoo2019photorealistic}(SS) & 95.65 & 8.78 & 92.46 & 74.11 & 25.53 & 4.55 \\
    \hline
    \textbf{WaveCT-AIN(SS)} & 96.83 & \textbf{6.91} & 93.99 & 78.63 & 28.34 & \textbf{0.08} \\
    \textbf{WaveCT-AIN-D(fix)} & 96.12 & 8.72 & 93.54 & 79.25 & 29.35 & 0.10 \\
    \textbf{WaveCT-AIN-D(SS)} &\textbf{97.31} & 7.54 & \textbf{94.50} & \textbf{79.65} & \textbf{30.44} & 0.11 \\
    \hline
    \end{tabular}%
  \label{tab:comparison_hc}%
\end{table}%

\textbf{Runtime comparison.} The last columns in Table \ref{tab:comparison_ac} and \ref{tab:comparison_hc} show the runtime comparisons among different methods using a single GTX 1080 Ti. We used the same hyperparameter settings in \cite{ma2019neural} for Gatys \textit{et al.} \cite{gatys2016image} and followed the original settings in STROTSS \cite{kolkin2019style}. The online style transfer methods above require much time to iteratively reconstruct a fluctuating stylized result, which goes against the clinical requirements. Because AdaIN \cite{huang2017arbitrary} is much simpler than WCT \cite{li2017universal}, the proposed WaveCT-AIN has improved the speed around 60 times when compared to WCT$^2$ \cite{yoo2019photorealistic}. \par

\textbf{Visual quality comparison.} To evaluate the visual quality, we calculated the structural similarity (SSIM) and Peak Signal to Noise Ratio (PSNR) between original images and stylized results. Based on the SSIM and PSNR in Table \ref{tab:comparison_ac} and \ref{tab:comparison_hc}, we can see that the visual quality of our method is superior to other methods. Moreover, in Gatys \textit{et al.} \cite{gatys2016image}, the visual quality of stylized results with average style transfer (fix) is much worse than the results based on style selection (SS) method. In contrast, the proposed WaveCT-AIN-D is robust to the style change. \par

\begin{figure}[htb]
	\centering
	\includegraphics[width=1\linewidth]{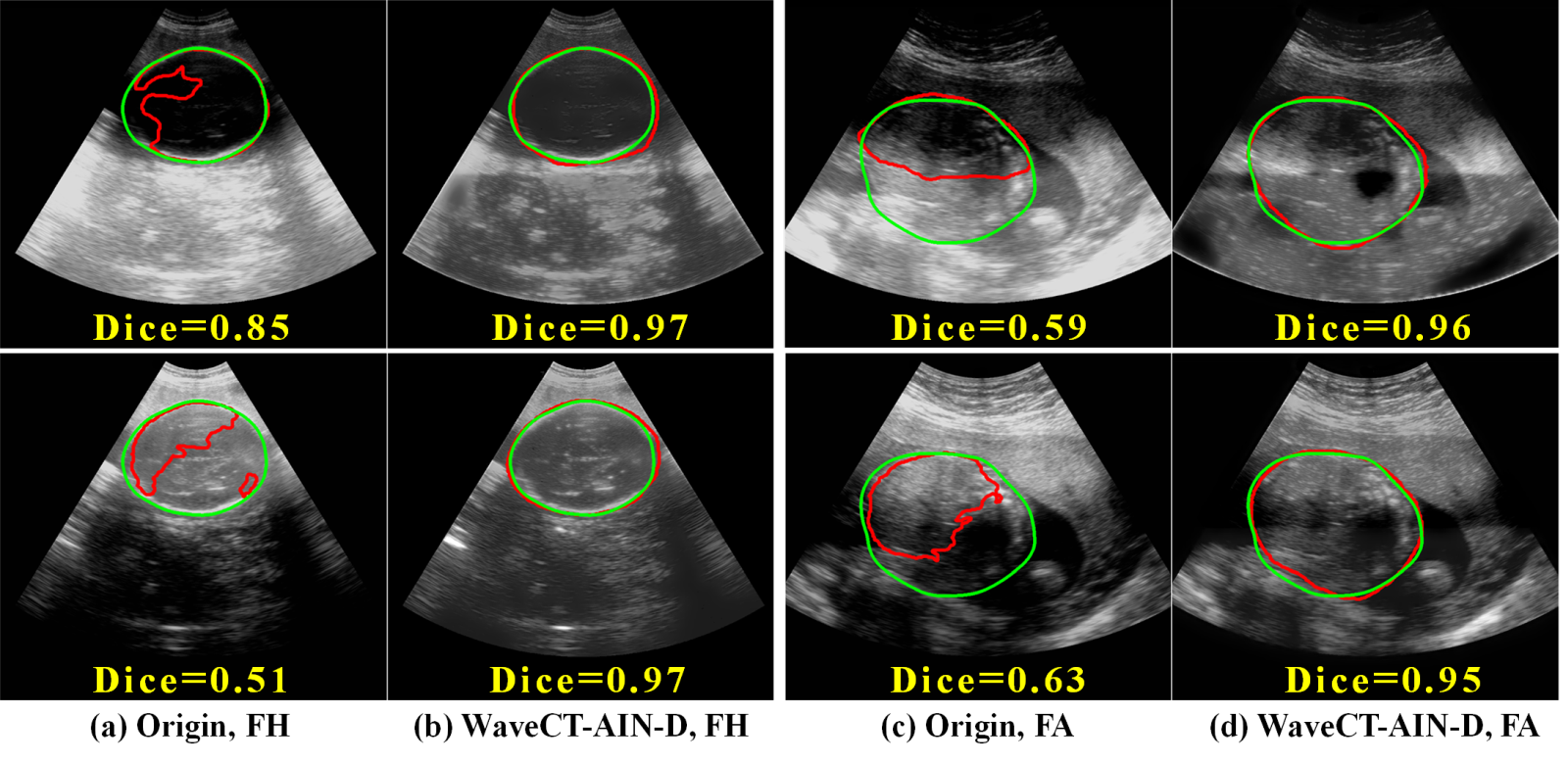}
	\caption{Examples of segmentation results using WaveCT-AIN-D. (a) and (c) are FH and FA ultrasound images with different TGC, respectively, while (b) and (d) are corresponding WaveCT-AIN-D results.}
	\label{fig:performance}
\end{figure}

\textbf{Segmentation performance comparison.} As shown in Table \ref{tab:comparison_ac} and \ref{tab:comparison_hc}, histogram equalization (HE) still suffer from appearance shift, but the proposed WaveCT-AIN-D (SS) obtain the highest Dice of 94.05\% and 97.31\% for FA and FH, respectively, which are also higher than without processing (91.0\%). Compared to other methods, our proposed methods have also achieved significant improvement in terms of Hausdorff distance boundary (Hdb) and Jaccard. Fig.\ref{fig:performance} demonstrates the effectiveness of WaveCT-AIN-D. \par

\label{sec:typestyle}

\section{Conclusions}

In this paper, we proposed a novel style transfer framework, called WaveCT-AIN-D, to eliminate image appearance shift in US image segmentation. The proposed method is faster, more stable and efficient than the state-of-the-art methods. In addition, we explored a style selection method to match style-content pairs appropriately. Future study may focus on validating the proposed framework on other modalities and clinical practice.

\bibliographystyle{IEEEbib}
\bibliography{refs}

\end{document}